\documentstyle[12pt,epsf]{article}
\begin{document}

\begin{flushright}
preprint DO-TH-95/10 \\
hep-ph/9507285 \\
July 11th 1995 \\ [1.5cm]
\end{flushright}

\begin{center}
{\Large \bf A dynamical model for nuclear structure functions} \\ [1cm]
{\bf D.\ Indumathi and W.\ Zhu}\footnote{On
leave from the Department of Physics, East China Normal University,
Shanghai 200062, P.\ R.\ China} \\
\normalsize Institut f\"ur Physik, Universit\"at Dortmund, Germany
\end{center}

\vspace{0.2cm}

\begin{abstract}
\noindent The dynamical origins of the EMC effect are studied. We
conclude that a
swelling in size of a bound nucleon as well as nuclear binding plays an
important r\^ole in determining the parton distributions within a bound
nucleon. We find that the distortion of nucleon structure functions
in nuclei can be simply explained with a few fundamental nuclear
parameters.

\vspace{0.3cm}


\end{abstract}

\vspace{0.5cm}


The fact that the structure functions of bound and free
nucleons are
not equal is called the EMC effect \cite{EMC1}. Recent accurate data
\cite{EMC2,data1,data2} on nuclear structure functions impels us
to reconsider the origin of the EMC effect. In this letter we report
that this effect can be explained in a broad kinematical region using the
idea of swelling and incorporating binding effects using only a
few fundamental nuclear parameters but with a new understanding
concerning these concepts. We expound on our ideas as follows.

As the first step, we choose a set of parton distributions which
can describe the structure functions of the free nucleon in the
kinematic region of the EMC effect (where $Q^2$ ranges from less than
$1~{\rm GeV}^2$ to a hundred ${\rm GeV}^2$). One such model was proposed
by Gl\"uck, Reya, and Vogt \cite{GRV}. The model assumes that there
exists a scale $Q^2=\mu^2$ which separates the perturbative regime
($Q^2\geq\mu^2$) from the non-perturbative one ($Q^2<\Lambda^2 $).
All parton distributions are generated dynamically by evolution from a
set of valence-like inputs (of the form
$N x^\alpha P_{N, q}(x) (1-x)^\beta$) at $\mu^2$.
For example, in the leading order (LO) approximation with $\mu^2=0.23\,
{\rm GeV}^2$, the valence ($u_v$, $d_v$), the total sea ($S$), and
gluon ($g$) input distributions are given to be \cite{GRV},
$$
\begin{array}{rcl}
xu_v^N(x) & = & 1.377x^{0.549}P_{N, u} (x)(1-x)^{3.027}~, \\
xd_v^N(x) & = & 0.328x^{0.366}P_{N, d} (x)(1-x)^{3.744}~, \\
xS_N(x) & = & 2 x(\overline{u}+\overline{d})
			    = 2.40x^{0.29}P_{N, S} (x)(1-x)^{7.88}~, \\
xg_N(x,) & = & 35.8x^{2.3}(1-x)^{4.0}~,
\end{array}
\eqno(1a)
$$
where
$$
\begin{array}{rcl}
P_{N, u} (x) & = & 1+0.81\sqrt{x}-4.36x+19.4x^{3/2}, \\
P_{N, d} (x) & = & 1+1.14\sqrt{x}+5.71x+16.9x^{3/2}, \\
P_{N, S} (x) & = & 1+0.31x~.
\end{array}
\eqno(1b)
$$
The shape of the valence quark distributions in (1) is similar to that of
the gluons, while the input sea quarks are much softer. This can
be understood if we assume that the gluons are co-moving with the
valence quarks whereas the sea quarks are mesonic at
$Q^2=\mu^2$, i.e., the input sea quarks correspond to
the mesonic component of the nucleon at $Q^2<\Lambda^2$, so that their
distribution will be softened by a convolution form \cite{Sull}.

We now discuss the nuclear binding effect at
$Q^2=\mu^2$.
Fermi motion is a global property of the bound nucleon. We regard the
energy of Fermi motion as
a part of the effective mass of the bound nucleon. This does
not change the total momentum of the nucleon but only smears the parton
distributions in the large $x$ region. Since we are interested here
primarily in the small and intermediate $x$ region, we do not consider this
effect in what follows. Therefore, the momentum of the struck nucleon
in the lab frame is
$P_N=(M_N-2b, \bf{p})$, where $b$ is the binding energy per nucleon.
The factor 2 is due to the fact that the nucleus with mass $(A-1)$ is
required to be on mass-shell. Our
approach is different from traditional models of the EMC effect
\cite{LS} since we do not need to compensate for the binding energy.
Instead, we try by using the Weizs\"acker mass formula \cite{Weiz} to
establish the connection between the binding effect and parton
distributions in nuclei. According to this formula, the
binding energy per nucleon arising strictly from the nuclear force is
$$
b=[1-P_s(A)]a_{\rm vol}+P_s(A)\frac{a_{\rm vol}}{2}
   = a_{\rm vol}-a_{\rm sur}A^{-1/3}~,
\eqno(2)
$$
for $A > 12$
where $P_s(A)$ is the probability of finding a nucleon on the nuclear
surface. We have ignored other contributions, especially the Coulombic
one, which is {\it not} probed in DIS.
Experimentally $a_{\rm vol}=15.67$ MeV, and $a_{\rm sur}=17.23$ MeV.

On the other hand, the attractive potential
describing the nuclear force is from the exchange of scalar mesons.
The energy required for this nuclear binding is taken away from individual
nucleons which thus lose energy.
We assume that the binding energy, $b$, corresponds
to loss of energy of ``mesonic'' sea quarks in the nucleons.
This means that
the momentum fraction carried by the sea quarks in a nucleon bound in a
nucleus at $Q^2=\mu^2$ will be reduced to
$$
\begin{array}{rcl}
S_A (x, \mu^2) & = & K(A)S_N(x,\mu^2) \\
 & = & \left(1-{\displaystyle 2b \over \displaystyle M_N\langle
        S_N(\mu^2)\rangle_2}\right) S_N(x,\mu^2)~.
\end{array}
\eqno(3)
$$
Here $\langle S_N\rangle_2$ is the momentum fraction (second moment) of
the sea quarks and we assume that the decrease in number of sea quarks
due to the binding effect is proportional to the quark density.

A bound nucleon will physically swell in size and this has been
discussed in the so-called rescaling models, in which the rescaling of
the input
point, $\mu^2$, is assumed \cite{Jaffe}. However, we consider that the
swelling of the nucleon only geometrically redistributes partons inside
the nucleon and does not change either the value of this dynamical
parameter or the existing number of partons at $\mu^2$. We try to
describe the swelling effect using some universal principles as in
\cite{Zhu}. The relative increase in the nucleon's radius is $\delta_A$,
where $(R_N+\Delta R(A))/R_N = 1 + \delta_A$. Analogous to (2), we assume
that the swelling of a nucleon on the nuclear surface is less than
that of one in the interior; therefore,
$$
\delta_A=[1-P_s(A)]\delta_{\rm vol}+P_s(A)\delta_{\rm vol}/2~.
\eqno(4)
$$
Here $\delta_{\rm vol}$ parametrises the swelling of the nucleon in the
interior of a heavy nucleus and is a constant for nuclei with $A > 12$
and also for Helium (with $P_s = 1$) since they have similar nuclear
densities. Interestingly, the distortions of the density
distributions and hence
the changes in the three main parameters, $N$, $\alpha$, and $\beta$,
in (1), due to swelling (if we assume $P_{A, q}(x) = P_{N, q}(x)$),
can be simply determined by some universal
principles. These cause the first three moments of the parton
distributions in a free ($q_N$) and bound ($q_A$) nucleon (for $q=$
valence, sea quarks, and gluons) at $\mu^2$ to be related by
$$
\begin{array}{rcl}
\langle q_A(\mu^2)\rangle_1  & = & \langle q_N(\mu^2)\rangle_1~, \\
\langle q_A(\mu^2)\rangle_2  & = & \langle q_N(\mu^2)\rangle_2~, \\
{\displaystyle (\langle q_N(\mu^2)\rangle_3-\langle q_N(\mu^2)
     \rangle_2^2)^{1/2} \over \displaystyle (\langle q_A(\mu^2)
     \rangle_3-\langle q_A(\mu^2)\rangle_2^2)^{1/2}} & = & 1+\delta_A~.
\end{array}
\eqno(5)
$$
The first two equations imply number and momentum conservation of
partons and the last incorporates the swelling effect \cite{Zhu}.
Fig.\ 1 gives an example of the swelling effect for the parton densities
in calcium. The momenta lost from the small and large $x$ regions are
transferred to the intermediate $x$ region.
The effect thus weakly enhances the distributions in the region
$0.1 < x < 0.3$ and results in ``antishadowing''. Furthermore, this
enhancement
depends on the nuclear density and does not disappear at larger $Q^2$.

Finally, there is a further depletion of the sea densities {\it at the
time of} scattering. This is easiest to see in the Breit frame, where the
exchanged virtual boson is completely space-like, so that the
3--momentum of the struck parton is flipped in the interaction. Hence,
a struck parton carrying a fraction $x$ of the nucleon's momentum,
$P_N$, during the interaction time $\tau_{\rm int}=1/\nu$,
will be localized longitudinally to
within a potentially large distance $\Delta Z\sim 1/(2xP_N)$.
Although the spatial extent of a single colored parton cannot exceed
the range of QCD confinement, one can assume that a
struck parton with $x < x_0 = 1/(2 M_N d_N)$ can combine with a wee parton
and form a colorless state $T_0$ with vacuum quantum numbers. Here
$d_N$ is the average correlation distance between two neighboring nucleons
in the lab frame:
$$
\begin{array}{rcl}
d_N & = & P_s(A) \left[R_N+2 (R_{\rm WS}-R_N)\right] + \\
 & & \hspace{1cm} (1-P_s(A)) \left[{\displaystyle R_N \over
 \displaystyle 2} + 2 (R_{\rm WS}-R_N)\right]~,
\end{array}
\eqno(6)
$$
where $R_N$ and $R_{\rm WS}$ are the nucleon and Wigner-Staiz radii.
The observable value of the momentum of $T_0$ is its average over the
uncertainty time $\tau_{\rm int}$ and equals zero in the Breit frame.
Therefore, $T_0$ is a static scalar field.

\begin{center}
\leavevmode
\epsfxsize=3.5in
\epsfbox{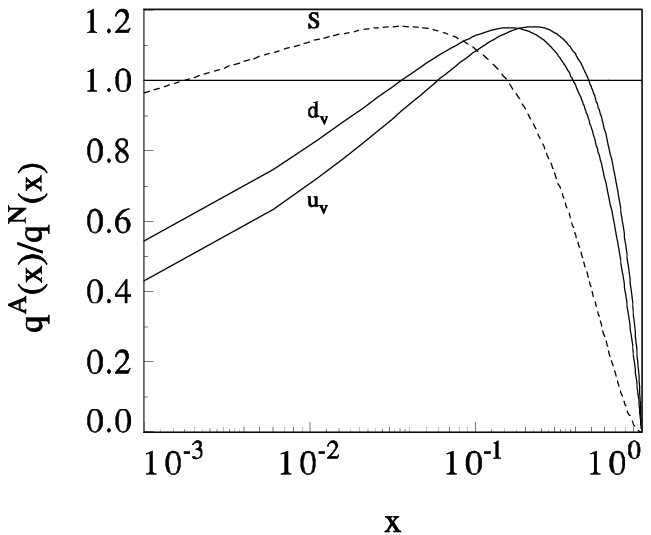}

\vspace{0.3cm}

\end{center}

\noindent {\sl Fig.\ 1  The effect of nucleon swelling on the calcium input
distributions: the ratios of the modified to unmodified densities are
shown for $u_v$, $d_v$, and $S$. }

\vspace{0.3cm}

We know that the correlation between two nucleons via scalar field
exchange is the bound state problem. Therefore, in the DIS process,
a nucleon can interact with other nucleons via $T_0$
in the long-distance.
Because of this, a new binding effect will
be experienced by the correlated nucleons, which we call the second
binding effect.

Consider a pair of nucleons which are correlated by $T_0$. The
interaction causes the number of sea quarks $S_A(x,Q^2)$ in a nucleon
to be further reduced to $S_A'(x,Q^2)$. (Valence quarks are not
depleted due to the requirement of quantum number conservation).
Similar to (3), one can simply assume that the loss in
number of sea quarks is
proportional to the original density, i.e., $S_A(x,Q^2)-S_A'(x,Q^2)=
\beta S_A(x,Q^2),$
where $\beta$ is a constant. The total energy loss in a bound nucleon
due to the second binding effect therefore is
$$
U_s(Q^2) = \beta M_N\int_{0}^{x_0}xS_A(x,Q^2)
    \simeq\beta M_N\langle S_A(Q^2)\rangle_2~.
\eqno(7)
$$
Hence,
$$
\beta=\frac{U_s(Q^2)}{M_N\langle S_A(Q^2)\rangle_2}
       =\frac{U(\mu^2)}{M_N\langle S_N(\mu^2)\rangle_2}~,
\eqno(8)
$$
$U(\mu^2)=a_{\rm vol}/6$ being the binding energy between each pair
of nucleons.

Partons with momentum fraction $x$ can overlap $(n-1)$ other nucleons,
where $n=1/(2 M_N d_N x) = x_0/x$.  The second binding effect
arising from each overlap is simply additive due to the applicability of
the superposition principle to the linear static field. The
corresponding shadowing of the sea quarks is given by $S'_A(x, Q^2)
= K'(A) S_A(x, Q^2)$, where the depletion factor is,
$$
\begin{array}{rcll}
K'(A) & = & 1, & \hbox{when}~x > x_0; \\
& = & 1-2\beta(x_0x^{-1}-1), & \hbox{when}~x_A < x < x_0; \\
& = & 1-2\beta(x_0x_A^{-1}-1), & \hbox{when}~x < x_A,
\end{array}
\eqno(9)$$
where $2\beta = 0.037$, $x_A=1/(4\overline{R}_AM_N)$, and
$2 \overline{R}_A \simeq 1.4 R_A$
is the average thickness of the nucleus. We emphasize that the second
binding effect acts on the intermediate state of the probe--target
interaction and does not participate in the QCD evolution of the initial
state. Since $x_0 \,{\stackrel{\displaystyle {}_<}{\displaystyle {}_\sim}}
\,0.1$, this is also a small $x$ effect.

In summary, the input nucleon structure function
(for an isoscalar nucleus) is given by
$$
\begin{array}{rcl}
F_A(x,\mu^2) & = & \langle e^2\rangle \left[\rule{0mm}{4mm}
		    xu_v^A(x,\mu^2) + xd_v^A(x,\mu^2) \right. \\
  & & \left. \hspace{1cm} + K(A)xS_A(x,\mu^2) \rule{0mm}{4mm} \right]~,
\end{array}
\eqno(10)
$$
where the average charge square factor is $\langle  e^2\rangle = 5/18$
and $q^A(x,\mu^2)$ incorporates the effect of swelling on every
input parton density, $q^N(x,\mu^2)$.
The structure function ratio of nucleons bound in two different nuclei,
$A$ and $B$, at $Q^2>\mu^2$ is given by,
$$
R^{\rm AB}={\displaystyle xu_v^A(x,Q^2)+xd_v^A(x,Q^2)+
	 K'(A)x\widetilde{S}_A(x,Q^2) \over \displaystyle
      xu_v^B(x,Q^2)+xd_v^B(x,Q^2)+K'(B) x\widetilde{S}_B(x,Q^2)}~,
\eqno(11)
$$
where $\widetilde{S}_A(x, Q^2)$ corresponds to the evolution of
$K(A) S_A (x, \mu^2)$, i.e., of the swelled input sea density, depleted
by the binding effect, to the scale $Q^2$. Note that for
$Q^2 > \mu^2$, the sea density also includes a finite contribution from
strange quarks.
The smearing effect of Fermi motion (at large $x$) is neglected in
this work.

In fig.\ 2 we give our predictions for the
$x$, $Q^2$, and $A$ dependences of the ratios for He/D, C/D and
Ca/D. The only free parameter to be fixed in computing these ratios is
the value of $\delta_{\rm vol}$. We find that $\delta_{\rm vol} = 0.15$
best describes the various available data for these nuclei.

We have also
shown the ratio of the ${}^6$Li to D structure functions in fig.\ 2.
Unlike helium (which also has $A < 12$), lithium is a very loosely bound
nucleus (whose density is less than half that of helium or carbon, while
its radius is larger than that of carbon \cite{EMC2}). Hence, $\delta_A$
for Li can be expected to be much smaller than that for He
($\delta_{\rm He} = 0.15/2$ according to (4)) and we take it to be 0.033.
In all cases, the swelling effect dominates the intermediate $x$
ratios while both the binding effects determine the small $x$ ratios.

The ratios $R^{\rm AB} = F^A_2/F^B_2$ for $A/B$ corresponding to
C/Li, Ca/Li, and Ca/C are more sensitively dependent on the
model of the EMC effect (as seen in fig.\ 3).
We predict, in general, an enhancement of the ratio, $R^{\rm AB}$, at
intermediate $x$ which is a net effect of the swelling and the first
binding effect, both of which depend on the nuclear density.
On the other hand, the strong
depletion at small $x$ depends on both the nuclear radius and density.
The weaker enhancement of the intermediate $x$ ratios of
Ca and C is because of their similar densities \cite{rad}
as compared to the C/Li and Ca/Li ratios where the two nuclei have
different densities. Our predictions are seen to be in excellent
agreement with the NMC data \cite{data1}, also shown for the
sake of comparison. Note that we have computed these three ratios at the
{\it same} values of $(x, Q^2)$ as the data.
We emphasize that, apart from the swelling parameter,
$\delta_{\rm vol}$, all parameters are fixed by a few fundamental
nuclear parameters.

\begin{center}
\leavevmode
\epsfxsize=5.5in
\epsfbox{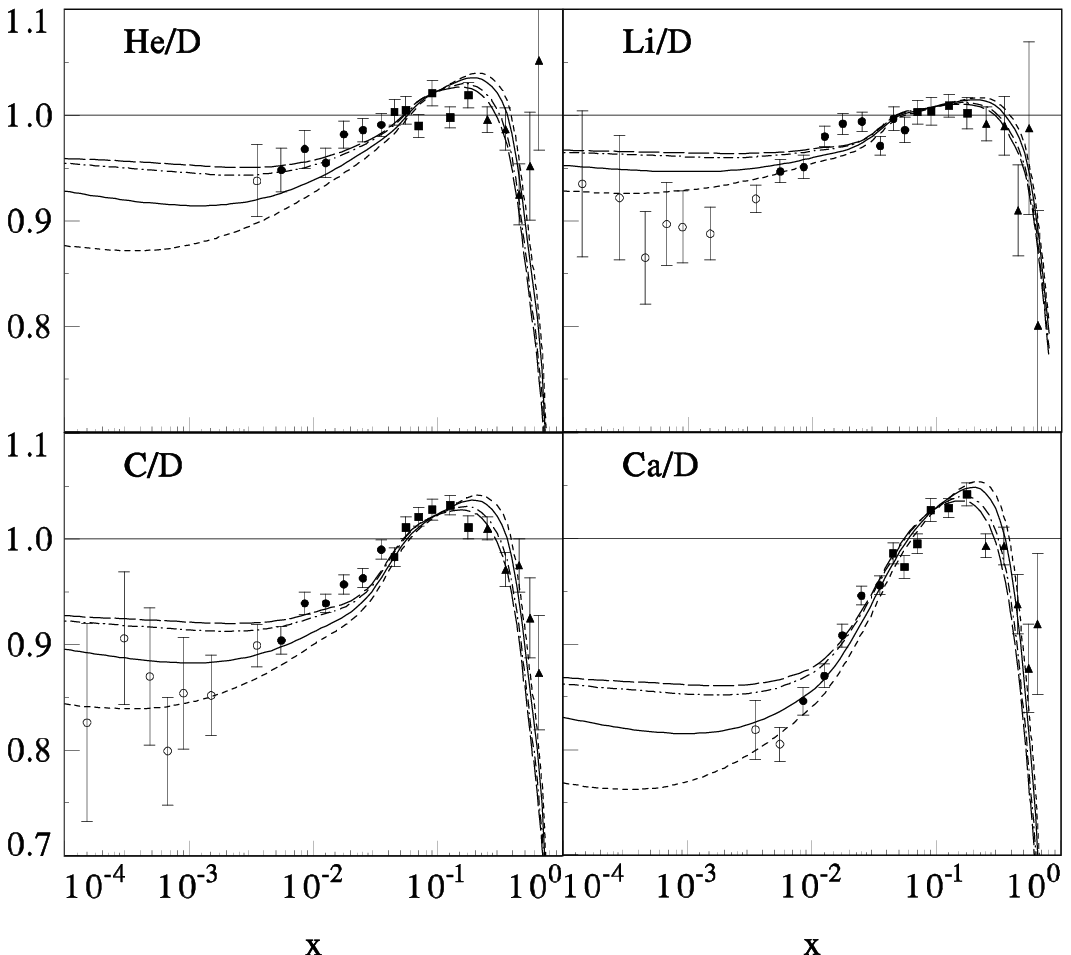}

\vspace{0.3cm}

\end{center}

\noindent {\sl Fig.\ 2 The structure function ratios as functions
of $x$ for (a) He/D, (b) Li/D, (c) C/D, and (d) Ca/D. The dashed, full,
broken, and long-dashed curves correspond to $Q^2 =$ 0.5, 1, 5, and
15~${\rm GeV}^2$ respectively. The data, \cite{data1,data2}, shown as open
and solid circles, boxes and triangles correspond to $Q^2 <$ 1,
1--5, 5--15, and $> 15~{\rm GeV}^2$ respectively.}

\vspace{0.3cm}

The input (1) reliably predicts the free nucleon stucture function in
the region $Q^2 > 0.5\,{\rm GeV}^2$.
Hence our dynamical model is also expected to predict correctly, not
only the ratios, but also the values of the nuclear structure
functions themselves in the same kinematic region.
The detailed comparison between the theory and the data
will published elsewhere.

In conclusion we have shown that the EMC effect can be explained
using swelling and binding effects but with a new understanding of these
ideas. As the general properties of the nuclear
force originated from the binding effect of nuclei in the history of
nuclear physics, we expect that a new understanding of binding effects
in the EMC effect will bring to light the nature of the nuclear force
at the level of quarks and gluons.

\begin{center}
\leavevmode
\epsfxsize=3.5in
\epsfbox{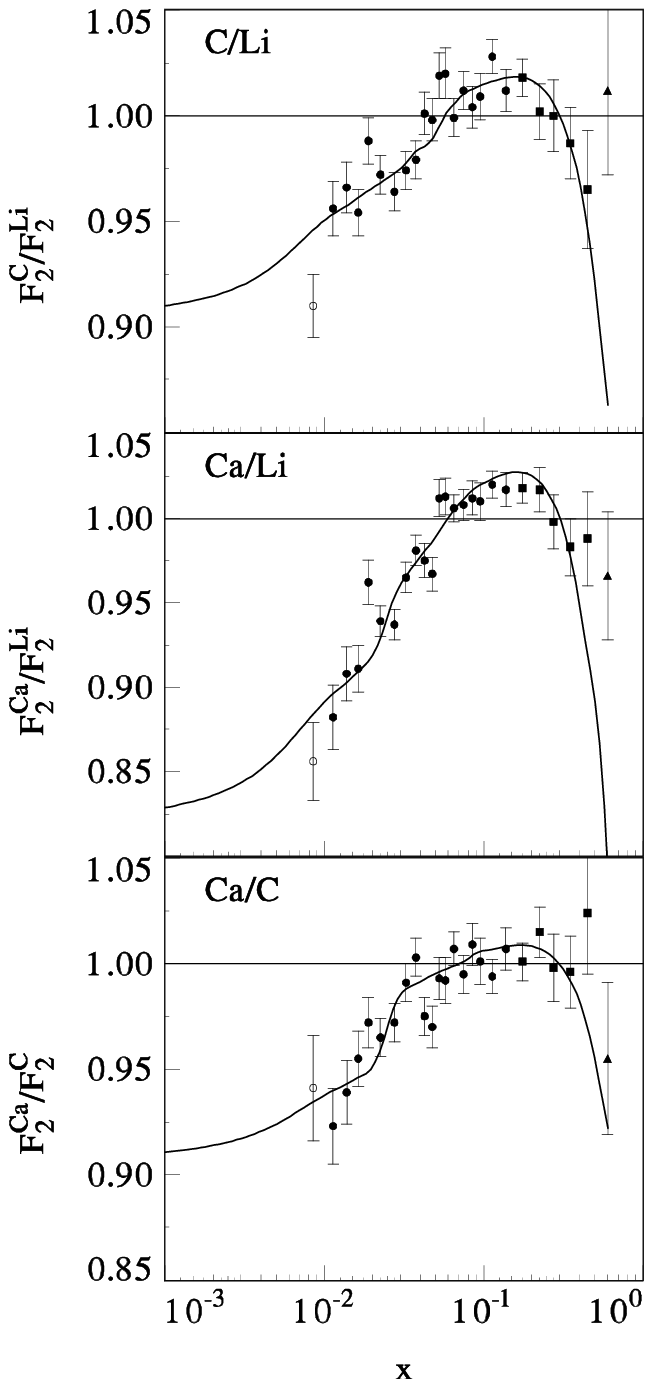}

\vspace{0.3cm}

\end{center}

\noindent {\sl Fig.\ 3  The structure function ratios for (a) C/Li,
(b) Ca/Li, and (c) Ca/C. The solid curve is a smooth fit to our
theoretical predictions at the same $(x, Q^2)$ as each available data
point \cite{data1}. The description of the data is the same as in
fig.\ 2.}

\vspace{0.3cm}

We thank M.\ Gl\"uck and E.\ Reya for extensive discussions and
encouragement.
One of us (W.\ Z) acknowledges the support of the DAAD--K.\ C.\ Wong
Fellowships and the National Natural Science Foundation of China and
the hospitality of the University of Dortmund where this work was done.



\begin{thebibliography}{99}
\bibitem{rad} The radii and densities, $(R_A~ \hbox{fm},
\rho~ \hbox{fm}^{-3})$, of the various nuclei, as quoted in \cite{EMC2},
are (2.6, 0.04) for ${}^{6}$Li, (2.5, 0.09) for ${}^{12}$C,
and (3.5, 0.11) for ${}^{40}$Ca.
\end{thebibliography}
\end{document}